\documentclass[aps,prd,nofootinbib,superscriptaddress]{revtex4-2}
\usepackage{graphicx,multirow,subfig}
\usepackage{bm}
\usepackage{braket}
\usepackage{amsmath}
\usepackage{amssymb}
\usepackage{amscd}
\usepackage{latexsym}
\usepackage{slashed}
\usepackage{color}
\usepackage{graphicx}
\usepackage[normalem]{ulem}
\usepackage{color}
\definecolor{orange}{cmyk}{0,0.5,1,0}
\usepackage{verbatim}
\usepackage{rotating}
\usepackage{diagbox}
\usepackage{cancel}
\usepackage[utf8]{inputenc}
\usepackage{array}
\usepackage{hyperref}
\usepackage[normalem]{ulem}

\def\lsim{\raise0.3ex\hbox{$\;<$\kern-0.75em\raise-1.1ex\hbox{$\sim\;$}}}
\def\gsim{\raise0.3ex\hbox{$\;>$\kern-0.75em\raise-1.1ex\hbox{$\sim\;$}}}

\newcolumntype{L}[1]{>{\raggedright\let\newline\\\arraybackslash\hspace{0pt}}m{#1}}
\newcolumntype{C}[1]{>{\centering\let\newline\\\arraybackslash\hspace{0pt}}m{#1}}
\newcolumntype{R}[1]{>{\raggedleft\let\newline\\\arraybackslash\hspace{0pt}}m{#1}}

\def\be{\begin{equation}}
\def\ee{\end{equation}}
\def\bea{\begin{eqnarray}}
\def\eea{\end{eqnarray}}

\newcommand{\doublet}[2]{ \left( \begin{array}{c}#1 \\ #2 \end{array}\right) }

\newcommand{\xdownarrow}[1]{%
	{\left\downarrow\vbox to #1{}\right.\kern-\nulldelimiterspace}
}

\begin{document}
\title{The $Z_3$ symmetric I(2+1)HDMl}

\author{A.~Aranda}
\email[]{fefo@ucol.mx}
\affiliation{Facultad de Ciencias-CUICBAS, Universidad de Colima, Bernal D\'iaz del Castillo 340, Colima 28045, M\'exico}

\author{D.~Hern\'andez-Otero}
\email[]{danielah@ifuap.buap.mx}
\affiliation{\small Instituto de F\'isica, Benem\'erita Universidad Aut\'onoma de Puebla, Apdo. Postal J-48, C.P. 72570 Puebla, Puebla, M\'exico}
	
\author{J.~Hern\'andez-Sanchez}
\email[]{jaime.hernandez@correo.buap.mx}
\affiliation{\small Facultad de Ciencias de la Electr\'onica, Benem\'erita Universidad Aut\'onoma de Puebla, Apdo. Postal 542, C.P. 72570 Puebla, Puebla, M\'exico}
	
\author{V.~Keus}
\email[]{venus.keus@helsinki.fi}
\affiliation{\small Department of Physics and Helsinki Institute of Physics, Gustaf Hallstromin katu 2, FIN-00014 University of Helsinki, Finland}
\affiliation{\small School of Physics and Astronomy, University of Southampton,
	Southampton, SO17 1BJ, United Kingdom}

\author{S.~Moretti}
\email[]{S.Moretti@soton.ac.uk}
\affiliation{\small School of Physics and Astronomy, University of Southampton,
	Southampton, SO17 1BJ, United Kingdom}

\author{D.~Rojas-Ciofalo}
\email[]{D.Rojas-Ciofalo@soton.ac.uk}
\affiliation{\small School of Physics and Astronomy, University of Southampton,
	Southampton, SO17 1BJ, United Kingdom}
	
\author{T.~Shindou}
\email[]{shindou@cc.kogakuin.ac.jp}
\affiliation{Division of Liberal-Arts, Kogakuin University, 2665-1 Nakano-machi, Hachioji, Tokyo, 192-0015, Japan}
	
\begin{abstract}
We introduce a 3-Higgs Doublet Model (3HDM) with two Inert (or dark) scalar doublets and an active Higgs one, hence termed I(2+1)HDM, in the presence of a discrete $Z_3$ symmetry acting upon the three doublet fields. 
We show that such a construct yields a Dark Matter (DM) sector with two mass-degenerate states of opposite CP quantum number, both of which contribute to DM dynamics, which we 
call ``{Hermaphrodite DM}'', distinguishable from a (single) complex DM candidate.
We show that the relic density contributions of both states are equal, saturating the observed relic density compliant with (in)direct searches for DM as well as other experimental data impinging on both the dark and Higgs sectors of the model, chiefly, in the form of Electro-Weak Precision Observables, Standard Model-like Higgs boson measurements at the Large Hadron Collider and void searches for additional (pseudo)scalar states at the CERN machine and previous colliders.
\end{abstract}
\maketitle

\vspace*{0.5cm}
\section{Introduction}

The discovery of a Higgs boson by the Large Hadron Collider (LHC) in July 2012 \cite{Aad:2012tfa,Chatrchyan:2012ufa} has finally confirmed that Electro-Weak Symmetry Breaking (EWSB) is triggered by the Higgs mechanism. 
While such a new state of Nature is perfectly consistent with the Standard Model (SM), which incorporates one Higgs doublet, 
there is no compelling reason to assume that there should be only one. In fact, it is possible that additional Higgs doublets exist such that their corresponding Higgs bosons could  be found during one of the upcoming LHC runs.
If one assumes that doublet (complex) representations of  Higgs fields are those chosen by Nature to implement EWSB, which is entirely plausible in the light of the fact that only such a structure is able to give mass to the $W^\pm$ and $Z$ bosons of the SM while preserving a massless photon, thereby in turn enabling unification of Electro-Magnetic (EM) and weak interactions, then one may wonder what can  models with a generic number $N$ of Higgs doublets, in turn defining the class of $N$-Higgs Doublet Models (NHDMs), produce in terms of new physics signals. The question is particularly intriguing if one further connects it to the need to explain the existence of Dark Matter (DM) in Nature, something that is absent in the SM. 

In order to attempt answering such a more articulate question, one may concentrate on the class of 2-Higgs Doublet Models (2HDMs) \cite{Branco:2011iw}.  In doing so, one should make sure to realise a 2HDM in a structure within which the DM candidate is a \textit{stable}
(on cosmological time scales), \textit{cold} (i.e., non-relativistic) at the onset of galaxy formation, \textit{non-baryonic}, \textit{neutral} and \textit{weakly interacting} component of the Universe \cite{Ade:2013zuv}. A very simple 2HDM realisation that provides a scalar DM candidate is the model with 1 Inert (I) doublet plus 1 Higgs (H) doublet, that we
label as  I(1+1)HDM. This  2HDM representation is known in the literature as the Inert Doublet Model (IDM), which was proposed in 1978 \cite{Deshpande:1977rw} and  has been studied extensively over many decades. In this scenario,  one $SU(2)_L$ doublet with the same SM quantum numbers as the SM Higgs one is introduced. 
Here, a $Z_2$ symmetry is also introduced, under which the even parity is assigned to the SM Higgs doublet and the odd parity is assigned to the additional one.
A possible vacuum configuration of this model is $(v,0)$, where the second doublet does not develop a Vacuum Expectation Value (VEV) while the first one inherits the SM VEV, $v$.\footnote{The doublet that acquires a VEV is  called the \textit{active} doublet and the one with no VEV is  called the \textit{inert} (at times also {\it dark}) doublet.} 
With this vacuum configuration, the $Z_2$ symmetry remains un-broken after  EWSB. In fact, 
the former does not take part in EWSB while the latter contains essentially the aforementioned Higgs state discovered at the CERN machine. Since the inert doublet does not couple to fermions, as it is by construction the only $Z_2$-odd field in the model,  it provides a stable DM candidate.
In essence, this is the lightest state among the two neutral (scalar and pseudoscalar) inert states with  $Z_2$-odd quantum numbers (while all the SM states are $Z_2$-even).\footnote{Incidentally, notice that scalar $(H)$ and pseudoscalar $(A)$ particles from the inert doublet in the I(1+1)HDM have opposite CP quantum numbers but, as they do not couple to fermions, the only means of disentangling them is to exploit their  gauge interactions: e.g., the $HAZ$ vertex is present while the $HHZ$ and $AAZ$ ones are not.}

The next class of NHDMs is constituted by 3-Higgs Doublet Models (3HDMs). The case for these is particularly promising for two main reasons. 
To begin with, 3HDMs are more tractable than higher multiplicity  NHDMs as all possible finite symmetries have been identified \cite{Ivanov:2012fp,Darvishi:2019dbh}.
Furthermore, and perhaps most intriguingly, 3HDMs may shed light on the flavour problem, namely the problem of the origin and nature of the three families of quarks and leptons, including neutrinos, and their pattern of masses, mixings and CP violation. Indeed, it is possible that the three families of SM fermions  could be described by the same symmetries that describe the three Higgs doublets \cite{Weinberg:1976hu}. 
In such models this family symmetry could be spontaneously broken along with the EW one, with some remnant subgroup  surviving, so that, for
 certain symmetries, it is possible to find a VEV alignment that respects the original symmetry of the scalar potential which will then be responsible for the stabilisation of the DM candidate \cite{Ivanov:2012hc}.

One could then simply extend the I(1+1)HDM by introducing another \textit{inert} $SU(2)_L$ doublet with, again,  the same SM quantum numbers as the SM Higgs one, thereby realising a
I(2+1)HDM, hence  achieving the vacuum alignment $(0,0,v)$, which is of particular interest because of its similarity with the I(1+1)HDM  and the absence of Flavour Changing Neutral Currents (FCNCs).\footnote{A 3HDM with $(0,v,v')$ vacuum alignment has been considered in \cite{Grzadkowski:2010au} wherein it was termed IDM2. Using our nomenclature, this model may be referred to as the I(1+2)HDM.} 

{The I(2+1)HDM has the following advantages over the I(1+1)HDM. Firstly, 
the low and medium mass regions for the DM candidate ($m_{\rm DM} \lesssim 100$ GeV) which are excluded in the I(1+1)HDM are revived in the I(2+1)HDM due to the presence of more coannihilation channels \cite{Keus:2014jha}. Secondly, the extended dark sector in the I(2+1)HDM allows for the possibility of \textit{dark CP violation}, a novel phenomenon first introduced in \cite{Cordero-Cid:2016krd}, which further revives the low DM mass region in comparison to the I(1+1)HDM. Note that, in order to introduce dark CP violation, it is necessary to at least add a singlet scalar to the I(1+1)HDM. However, the CP violating effects are smaller and less accessible to measure in the I(1+1)HDM plus a singlet compared to the I(2+1)HDM \cite{Cordero-Cid:2020yba}. Thirdly, in order 
to impose a $Z_N$ symmetry with $N>2$ on the dark sector, again, at least one singlet should be added to the I(1+1)HDM. However, in this paper, we will probe $Z_3$ symmetric I(2+1)IDM solutions that satisfy several constraints for the low DM mass region which are insteady excluded in the $Z_3$ symmetric I(1+1)HDM plus singlet.}
This $Z_3$ symmetric I(2+1)HDM is the model we will be concerned with, building upon the one introduced and studied in Refs. \cite{Cordero-Cid:2018man}--\cite{Cordero:2017owj}. Herein, though, the discrete symmetry structure used was again a $Z_2$ one, like in 2HDMs, separating the two inert doublets and the active one. Again, the lightest $Z_2$-odd neutral scalar of this construct is the DM candidate. 

In this paper, we study a variation of such a I(2+1)HDM, wherein we replace this $Z_2$ symmetry with a $Z_3$ one instead, following the example adopted in \cite{Aranda:2014lna} for the case of a 2HDM. The motivation for this is to attempt generating a 3HDM with two mass-degenerate DM states with opposite CP. We shall in fact show that, in the case of a $Z_3$ symmetric scalar potential, such a DM set-up, which we call ``{Hermaphrodite DM}'', is indeed possible and distinguishable from a complex DM set-up.

The layout of the remainder of the paper is as follows. In the next section we describe  the aforementioned variation of the I(2+1)HDM with a $Z_3$ symmetry. In the following sections we discuss both theoretical and experimental constraints affecting our
model. Numerical results and the selection of our benchmark scenarios will  then follow  while in the last  section we will conclude.

\section{The I(2+1)HDM scalar potential}
\label{3HDM}

In an NHDM, the generic scalar potential symmetric under a group $G$ of phase rotations can be written as the sum of two parts: 
\be 
V = V_0 + V_G,
\label{V0-3HDM}
\ee
where $V_0$ is invariant under any phase rotation and $V_G$ is a collection of extra terms ensuring the symmetry under the action of the group $G$ \cite{Ivanov:2011ae}.

The most general phase invariant part of a 3HDM potential has the following form:
\bea
V_0 &=& - \mu^2_{1} (\phi_1^\dagger \phi_1) -\mu^2_2 (\phi_2^\dagger \phi_2) - \mu^2_3(\phi_3^\dagger \phi_3) \\
&&+ \lambda_{11} (\phi_1^\dagger \phi_1)^2+ \lambda_{22} (\phi_2^\dagger \phi_2)^2  + \lambda_{33} (\phi_3^\dagger \phi_3)^2 \nonumber\\
&& + \lambda_{12}  (\phi_1^\dagger \phi_1)(\phi_2^\dagger \phi_2)
+ \lambda_{23}  (\phi_2^\dagger \phi_2)(\phi_3^\dagger \phi_3) + \lambda_{31} (\phi_3^\dagger \phi_3)(\phi_1^\dagger \phi_1) \nonumber\\
&& + \lambda'_{12} (\phi_1^\dagger \phi_2)(\phi_2^\dagger \phi_1) 
+ \lambda'_{23} (\phi_2^\dagger \phi_3)(\phi_3^\dagger \phi_2) + \lambda'_{31} (\phi_3^\dagger \phi_1)(\phi_1^\dagger \phi_3),  \nonumber
\eea
where the notation of \cite{Keus:2014jha} was used.
The construction of the $Z_3$ symmetric part of the potential depends on the generator of the $Z_3$ symmetry. As we want to study the model with two different DM candidates, in order to accomplish this, we will assign different charges to each doublet. 
More specifically, we assume that the Lagrangian is symmetric under the $Z_3$ transformation given by 
\begin{equation}
\phi_1^{}\to \omega \phi_1^{}\;,\quad 
\phi_2^{}\to \omega^2 \phi_2^{}\;, \quad 
\phi_3^{}\to \phi_3^{}\;,
\end{equation}
with $\omega$ being a complex cubic root of unity, $\omega = e^{2\pi i/3}$.
In other words, we can write the generator of the group as follows:
\be 
g_{Z_3}=  \mathrm{diag}\left(\omega, \omega^2, 1 \right).
\ee
With these assignments, 
the $Z_3$ symmetric potential term $V_G$ has 
the following form:

\be 
V_{Z_3} = \lambda_1(\phi_2^\dagger\phi_1)(\phi_3^\dagger\phi_1) + \lambda_2(\phi_1^\dagger\phi_2)(\phi_3^\dagger\phi_2) + \lambda_3(\phi_1^\dagger\phi_3)(\phi_2^\dagger\phi_3)  + h.c.
\label{Z_3-3HDM}
\ee
{We take all the parameters of the potential to be real.}
We will identify $\phi_3$ with the SM Higgs doublet and 
the $Z_3$ charges for all other SM particles are  considered to be zero.
The Yukawa Lagrangian in this model is identical to the SM Yukawa Lagrangian (with additional terms for right-handed neutrinos) given by 
\bea 
\mathcal{L}_{Y} &=& \Gamma^u_{mn} \bar{q}_{m,L} \tilde{\phi}_3 u_{n,R} + \Gamma^d_{mn} \bar{q}_{m,L} \phi_3 d_{n,R} \nonumber\\
&& +  \Gamma^e_{mn} \bar{l}_{m,L} \phi_3 e_{n,R} + \Gamma^{\nu}_{mn} \bar{l}_{m,L} \tilde{\phi}_3 {\nu}_{n,R} + h.c.  
\eea

We assume  the vacuum alignment 
 $\langle \phi_1\rangle =\langle \phi_2\rangle =0$ and 
$\langle \phi_3\rangle \neq 0$, so that the $Z_3$ symmetry is unbroken when  EWSB occurs via the Higgs mechanism.

\subsection{The mass eigenstates}
\label{no-Z3-subsection}
We define the components of each doublet as 
\be 
\phi_1= \doublet{$\begin{scriptsize}$ H^{0+}_1 $\end{scriptsize}$}{\frac{H^0_1+iA^0_1}{\sqrt{2}}} ,\quad 
\phi_2= \doublet{$\begin{scriptsize}$ H^{0+}_2 $\end{scriptsize}$}{\frac{H^0_2+iA^0_2}{\sqrt{2}}} , \quad 
\phi_3= \doublet{$\begin{scriptsize}$ H^{0+}_3 $\end{scriptsize}$}{\frac{v+H^0_3+iA^0_3}{\sqrt{2}}}.
\label{explicit-fields}
\ee

The vacuum condition that the point $(\phi_1^0, \phi_2^0,\phi_3^0)= (0,0,\frac{v}{\sqrt{2}})$ 
becomes the minimum of the potential leads to the relation 
\be
v^2= \frac{\mu^2_3}{\lambda_{33}}.
\ee
Expanding the potential around this vacuum point results in the mass spectrum below, where the pairs of scalar/pseudoscalar base fields $(H_{1,2}^{0}/A_{1,2}^{0})$ from  the inert doublets in Eq.~(\ref{explicit-fields}) are rotated by:
\be 
R_{\theta_i}= 
\left( \begin{array}{cc}
	\cos \theta_i & \sin \theta_i \\
	-\sin \theta_i & \cos \theta_i\\
\end{array} \right), 
\ee
where $\theta_i=\theta_{h},\theta_{a}$ are the rotation angles for the scalar and pseudoscalar matrices, respectively while there is no mixing between the charged states. The mass spectrum of all spin-0 particles of the I(2+1)HDM is presented bellow.\\

\noindent
\textbf{CP-even scalars:}
\bea
&& \textbf{h} : \quad m^2_{h}= 2\mu_3^2 = 2 \lambda_{33} v^2\\[1mm]
&& \textbf{H}_1 = \cos\theta_h H^0_{1}+ \sin\theta_h H^0_{2}  
\nonumber\\
&& \hspace{1cm}	
m^2_{H_1}=  \cos^2\theta_h(-\mu^2_1 + \Lambda_{1}) +\sin^2\theta_h (- \mu^2_2 + \Lambda_{2})  +  \sin\theta_h \cos\theta_h \, \lambda_3 v^2
\nonumber\\[1mm]
&& \textbf{H}_2 = -\sin\theta_h H^0_{1}+ \cos\theta_h H^0_{2} 
\nonumber\\
&& \hspace{1cm}	
m^2_{H_2}=  \sin^2\theta_h(-\mu^2_1 + \Lambda_{1}) +\cos^2\theta_h (- \mu^2_2 + \Lambda_{2})  - \sin\theta_h \cos\theta_h \, \lambda_3 v^2 
\nonumber\\[1mm]
&& \mbox{with} \quad \Lambda_{1}= \frac{1}{2}(\lambda_{31} + \lambda'_{31})v^2,  
\quad \Lambda_{2}= \frac{1}{2}(\lambda_{23} + \lambda'_{23})v^2,  
\quad \tan 2\theta_h = \frac{ - \lambda_3 v^2}{\mu^2_1 - \Lambda_1 - \mu^2_2 + \Lambda_2} 
\nonumber
\eea
\noindent
\textbf{CP-odd scalars:}
\bea
&& \textbf{A}_1 = \cos\theta_a A^0_{1} + \sin\theta_a A^0_{2}
\\
&& \hspace{1cm}	
m^2_{A_1}= \cos^2\theta_a(-\mu^2_1 + \Lambda_{1}) +\sin^2\theta_a  (- \mu^2_2 + \Lambda_{2}) - \sin\theta_a \cos\theta_a   \lambda_3 v^2
\nonumber\\[1mm]
&& \textbf{A}_2 = -\sin\theta_a A^0_{1} + \cos\theta_a A^0_{2}
\nonumber\\
&& \hspace{1cm}		
m^2_{A_2}= \sin^2\theta_a(-\mu^2_1 + \Lambda_{1}) + \cos^2\theta_a(- \mu^2_2 + \Lambda_{2})  +  \sin\theta_a \cos\theta_a \lambda_3 v^2
\nonumber\\[1mm]
&& \mbox{with} 
\qquad
\tan 2\theta_a = \frac{ \lambda_3 v^2}{\mu^2_1 - \Lambda_1 - \mu^2_2 + \Lambda_2}  \nonumber
\eea
\noindent
\textbf{Charged scalars:}
\bea
\label{charged-masses}
&& \textbf{H}^\pm_1 = H^{0\pm}_{1} , \qquad
m^2_{H^\pm_1}=-\mu^2_1 + \frac{1}{2}\lambda_{31}v^2
\\[1mm]
&& \textbf{H}^\pm_2 = H^{0\pm}_{2}, \qquad 
m^2_{H^\pm_2}=  - \mu^2_2 + \frac{1}{2}\lambda_{23}v^2   \nonumber
\eea
Note that $\tan\theta_a = - \tan\theta_h$ and the CP-even and CP-odd mass eigenstates can be written as 
\be 
\label{mass-gauge-Z3}
\left\{\begin{array}{c}
\textbf{H}_1 \equiv \phantom{-}\cos\theta_h \, H^0_{1}+ \sin\theta_h \, H^0_{2}  \\[2mm]
\textbf{H}_2 = -\sin\theta_h \, H^0_{1}+ \cos\theta_h \, H^0_{2} 
\end{array}
\right. 
~~~ \mbox{and} ~~~
\left\{\begin{array}{c}
\textbf{A}_1 \equiv \cos\theta_h \, A^0_{1}- \sin\theta_h \, A^0_{2}  \\[2mm]
\textbf{A}_2 = \sin\theta_h \, A^0_{1}+ \cos\theta_h \, A^0_{2} 
\end{array}
\right.
\ee
with masses
\bea
\label{mass-eigenstates-Z3}
&& m^2_{H_1}=
m^2_{A_1} = \cos^2\theta_h(-\mu^2_1 + \Lambda_{1}) +\sin^2\theta_h (- \mu^2_2 + \Lambda_{2})  +\sin\theta_h \cos\theta_h \lambda_3 v^2   \\[1mm]
&& m^2_{H_2}=
m^2_{A_2}= \sin^2\theta_h (-\mu^2_1 + \Lambda_{1}) +\cos^2\theta_h (- \mu^2_2 + \Lambda_{2})  - \sin\theta_h \cos\theta_h  \lambda_3 v^2 
\nonumber
\eea
Note that the degenerate fields $H_1$ and $A_1$ can be grouped together into a complex neutral field $N_1 = (H_1 + iA_1 )/\sqrt{2}$ (and $H_2$ and $A_2$ states into $N_2 = (H_2 + iA_2 )/\sqrt{2}$, correspondingly). When $\lambda_3=0$, the inert doublets decouple from each other and the complex fields $N_1$ and $N_2$ become eigenstates of the $Z_3$ symmetry, with the $Z_3$ charge $+1$ for $N_1$ and the $Z_3$ charge $-1$ for $N_2$.
In general, though, when $\lambda_3 \neq 0$,  the states $N_1$ and $N_2$ do not have defined $Z_3$ quantum numbers. We will discuss this further in section \ref{section-hermaphrodite} where we introduce the concept of Hermaphrodite' DM and discuss how it is distinguishable from a complex DM scenario.

We take the mass-degenerate $H_1$ and $A_1$ particles as constituents of the Hermaphrodite DM state, which are protected from decaying to SM particles through the unbroken $Z_3$ symmetry. Moreover, the only fields that transform trivially under the $Z_3$ symmetry are the SM fields and the fields from the only active scalar doublet, $\phi_3$, which plays the role of the SM Higgs doublet.

The (pseudo)scalar-gauge boson interaction plays an important role here, since a non-zero $H_1A_1Z$ vertex predicts a signal at direct detection experiments which contradicts the observation and rules out  the model as a viable DM framework. As we have show in Table \ref{vertex-table}, the $ZH_1A_1$ vertex is proportional to $\cos 2 \theta_h$. This vertex vanishes at the $\theta_h=\pi/4$ slice of the parameter space which is where we define our benchmark scenarios in the upcoming sections.

\begin{table}[t!]
\begin{center}
\begin{tabular}{|l|c|}\hline
vertex & vertex coefficient \\ \hline
\hline		
$Z H_i A_i$ 			 
&
$ \cos 2\theta_h  $ 
\\ 	
$Z H_i A_j$ &
$\sin 2\theta_h$ 
\\ 
$W^\pm H^\mp_i H_i$ & 
$\cos \theta_h$ 
\\ 
$W^\pm H^\mp_i H_j$ 		 
&
$\sin \theta_h$   
\\ 
$W^\pm H^\mp_i A_i$ 	 
&
$\cos\theta_h$ 
\\
$W^\pm H^\mp_i A_j$ 		 
&
$ \sin\theta_h$ 
\\
 \hline	
\end{tabular}
\caption{\small Angular dependence of (pseudo)scalar-gauge couplings in the $Z_3$ symmetric model.}
\label{vertex-table}
\end{center}
\end{table}

\subsection{Input parameters}
\label{input-params-section}

We write the parameters of the potential that are relevant for our numerical studies,
\be 
\mu_1^2, \; \mu_2^2, \; \lambda_{23}, \; \lambda_{31}, \; \lambda'_{23}, \; \lambda'_{31}, \; \lambda_3, \; \lambda_1, \; \lambda_2
\ee
in terms of physical quantities
\be 
 m_{H_1}, ~~  
m_{H_2},~~  
m_{H^\pm_1}, ~~  
m_{H^\pm_2}, ~~
\theta_h, ~~ g_{1}, ~~ g_{2}, ~~ \lambda_1, ~~ \lambda_2, 
\ee 
where $ 
g_{1} = g_{h H_1 H_1}/v$ and $g_{2} = g_{h H_1 H_2}/v$
and $\lambda_1$ and $\lambda_2$ which appear in the cubic inert scalar interactions. 
The conversion relations are as follows:
\begin{small}
\bea 
\lambda_3 
&=&
\frac{4 \sin (2 \theta_h) ({m^2_{H_2}}-{m^2_{H_1}})}{v^2
   (\cos (4 \theta_h)-3)},
\\[2mm]
\lambda'_{23} 
&=& 
\frac{2 \cos (2 \theta_h)
({m^2_{H_1}}-{m^2_{H_2}})}{v^2(\cos (4 \theta_h)-3)}
+
\frac{1}{v^2} \left( {m^2_{H_1}}+{m^2_{H_2}-2 {m^2_{H^\pm_2}}}
\right),
\nonumber
\\  [2mm] 
\lambda'_{31} &=&
\frac{2 \cos (2 \theta_h)
({m^2_{H_2}}-{m^2_{H_1}})}{v^2 (\cos (4 \theta_h)-3)}
+
\frac{1}{v^2} \left( {m^2_{H_1}}+{m^2_{H_2}}-2 {m^2_{H^\pm_1}} \right), 
\nonumber\\ 
\mu^2_1 
&=&
\frac{-1}{2(\cos (4 \theta_h)-3)}
\biggl( 
\cos (4 \theta_h) 
\left(-g_1  v^2+{m^2_{H_1}}+{m^2_{H_2}}\right)
+3 g_1  v^2
\nonumber\\
&&
+g_2 v^2 (-2 \sin (2 \theta_h)
+\sin (4  \theta_h)-2 \tan (\theta_h))
+2 \cos (2 \theta_h) ({m^2_{H_1}}-{m^2_{H_2}})
-7 {m^2_{H_1}}+{m^2_{H_2}}
\biggr),
\nonumber
\\[2mm]
\mu^2_2 &=&
\frac{-2}{4 (\cos (4 \theta_h)-3)}
\biggl(
\cos (4 \theta_h) 
\left(-g_1v^2+{m^2_{H_1}}+{m^2_{H_2}}\right)
+3 g_1 v^2
\nonumber\\
&&
+2g_2 v^2 \left(\cot (\theta_h)+4 \sin (\theta_h) \cos ^3(\theta_h)\right)
-7 {m^2_{H_1}}+{m^2_{H_2}}
-2 \cos (2 \theta_h) ({m^2_{H_1}}-{m^2_{H_2}})
\biggr),
\nonumber\\[2mm]
\lambda_{23} &=&
\frac{-1}{v^2 (\cos (4 \theta_h)-3)}
\biggl(
\cos (4 \theta_h) \left(-g_1v^2+{m^2_{H_1}}-2 {m^2_{H^\pm_2}}+{m^2_{H_2}}\right)
+3 g_1 v^2 +6{m^2_{H^\pm_2}}
\nonumber\\
&&
+2 g_2 v^2 \left(\cot (\theta_h)+4\sin (\theta_h) \cos ^3(\theta_h)\right)
+2 \cos (2\theta_h) ({m^2_{H_2}}-{m^2_{H_1}})-7 {m^2_{H_1}}
+{m^2_{H_2}}
\biggr),
\nonumber
\\[2mm]
\lambda_{31} &=&
\frac{1}{v^2 (\cos (4 \theta_h)-3)}
\biggl(
\cos (4 \theta_h)
\left(g_1 v^2+2{m^2_{H^\pm_1}}-{m^2_{H_1}}-{m^2_{H_2}}\right)
-3 g_1v^2 
-6 {m^2_{H^\pm_1}}
\nonumber\\
&&
+g_2 v^2 (2 \sin (2 \theta_h)-\sin (4\theta_h)+2 \tan (\theta_h))
+2 \cos(2 \theta_h) ({m^2_{H_2}}-{m^2_{H_1}})
+7{m^2_{H_1}}-{m^2_{H_2}}
\biggr).
\nonumber    
\eea
\end{small}

\section{Constraints on the model parameters}\label{constraints}
\label{sec:constraints}
As the third doublet is identified with the SM Higgs doublet, 
$\mu_3, \lambda_{33}$ are Higgs field parameters, renormalised by the Higgs mass. We use the value $m_h=125$~GeV for the latter, so that 
\begin{equation} 
 m^2_h = 2\mu^2_3 = 2\lambda_{33} v^2.
\end{equation}

{For the $V_0$ part of the potential to have a stable vacuum (bounded from below) \cite{Keus:2014isa,Faro:2019vcd}, the following conditions are required\footnote{These conditions emerge from requiring the quartic part of the potential to be positive as the fields $\phi_i \to \infty$. The ``copositivity" method suggested in \cite{Kannike:2012pe} will result in more restrictive constrains.}:
\bea
&& \bullet ~ \lambda_{11}, \,\lambda_{22}, \,\lambda_{33} \geq 0, \\[1mm]
&& \bullet ~ \widetilde{\lambda_{12}} \equiv \lambda_{12} + \lambda'_{12}  + \sqrt{\lambda_{11}\lambda_{22}} \geq 0, \nonumber\\[1mm]
&& \bullet ~ \widetilde{\lambda_{23}} \equiv\lambda_{23} + \lambda'_{23} + \sqrt{\lambda_{22}\lambda_{33}} \geq 0,\nonumber\\[1mm]
&& \bullet ~ \widetilde{\lambda_{31}} \equiv\lambda_{31} + \lambda'_{31} + \sqrt{\lambda_{33}\lambda_{11}} \geq 0,\nonumber\\[1mm]
&& \bullet ~ 
\sqrt{\lambda_{11}\lambda_{22}\lambda_{33}} + (\lambda_{12} + \lambda'_{12}) \sqrt{\lambda_{33}} 
+ (\lambda_{31} + \lambda'_{31}) \sqrt{\lambda_{22}} 
+ (\lambda_{23} + \lambda'_{23}) \sqrt{\lambda_{11}} 
+\sqrt{2 \, \widetilde{\lambda_{12}}\,\widetilde{\lambda_{31}}\,\widetilde{\lambda_{23}}} \geq 0.
\nonumber
\eea
We also require the parameters of the $V_{Z_3}$ part to be smaller than the parameters of the $V_0$ part:
\be 
\bullet ~ |\lambda_1|, |\lambda_2|, |\lambda_3| < |\lambda_{ii}|, |\lambda_{ij}|, |\lambda'_{ij}| , \quad i\neq j = 1,2,3,
\ee
so that they do not dominate the behaviour of the potential at high field values.
For the point $(0,0,\frac{v}{\sqrt{2}})$ to be the minimum of the potential, the mass eigenvalues must be positive. Therefore, it is required that:
\bea
&& \bullet ~ -\mu^2_1 + \lambda_{31} \frac{v^2}{2}  > 0, \\
&& \bullet ~ -\mu^2_2 + \lambda_{23} \frac{v^2}{2}  > 0, \nonumber\\
&& \bullet ~ -2(\mu^2_1+\mu^2_2)+v^2(\lambda_{23}+\lambda'_{23}+\lambda_{31}+\lambda'_{31})   >  \biggl|
2(\mu^2_1-\mu^2_2)+v^2(\lambda_{23}+\lambda'_{23}-\lambda_{31}-\lambda'_{31}) 
\biggr|,
\nonumber
\eea
which also renders the $(0,0,\frac{v}{\sqrt{2}})$  point the global minimum of the potential.
From the $S$-matrix unitarity for elastic scattering of 2-to-2 body bosonic states, the magnitude of combinations of the $\lambda$ parameters in the potential is constrained. In agreement with perturabivity bounds, we take the absolute values $|\lambda_i |\leq 2\pi$ which also satisfies the unitarity conditions in \cite{Bento:2017eti}.}

Measurements done at LEP limit the invisible decays of $Z$ and $W^\pm$ gauge bosons, requiring that  \cite{Cao:2007rm,Lundstrom:2008ai}
\bea 
&& \bullet ~ m_{H_i^\pm} + m_{H_i,A_i} > m_{W^\pm} \\
&&  \bullet ~ m_{H_i} + m_{A_i} > m_Z \nonumber\\
&&  \bullet ~ 2m_{H_i^\pm} > m_Z \nonumber
\eea

Also, LEP provides a model-independent lower limit for the mass of the charged scalars: 
\be 
 \bullet ~  m_{H^\pm_i} > 70-90 \quad \mbox{GeV}.
\ee
Searches for charginos and neutralinos at LEP have been translated into limits of region of masses in the I(1+1)HDM \cite{Lundstrom:2008ai} where for 
\be  
\bullet~
m_H < 80 \quad \mbox{GeV} \quad \mbox{and} \quad m_A < 100 \quad \mbox{GeV} \nonumber
\ee
the following region is excluded
\be 
 \bullet ~ m_A - m_H > 8 \quad \mbox{GeV}.  
\ee
We have taken this limit into account in our numerical studies for any pair of CP-even and CP-odd particles.

Finally, the decay width of the Higgs into a pair of the inert scalars with $m_{S_i} < m_h/2$ is
\be
\Gamma (h\to S_iS_j) = \frac{g^2_{hS_iS_j} v^2}{32\pi m_h^3}
\biggl[ \biggl(m_h^2-(m_{S_i}+m_{S_j})^2 \biggr)
\biggl(m_h^2-(m_{S_i}-m_{S_j})^2 \biggr)
\biggr]^{1/2},
\label{Eq:Gamma_inv}
\ee
with $S_i,S_j = H_1,A_1$ where $g_{hS_iS_j}\, v$ is the coefficient of the $hS_iS_j$ term in the Lagrangian and $m_{S_i}$ is the mass of the corresponding neutral inert particle. Experimental measurements of invisible Higgs decays limit models in which the Higgs boson can decay into lighter particles which escape detection. The current limits 
on the SM-like Higgs boson invisible (inv) Branching Ratio (BR) from the ATLAS experiment are \cite{ATLAS:2020kdi}
\be 
 \text{BR}(h\to \text{inv}) < 0.08-0.15.
\ee

This  leads to strong constraints on the {Higgs-DM} coupling.
For our scenarios this BR is: 
\be 
 \text{BR}(h\to \text{inv}) = \frac{\Sigma_{i}\Gamma(h\to S_iS_i)}{\Gamma^{\text{SM}}_h + \Sigma_{i}\Gamma(h\to S_iS_i)},
 \label{Eq:BRinv}
\ee
where $S_i =H_1, \, A_1$.

Regarding constraints coming from $h\to\gamma\gamma$, the inert charged masses and parameters in our analysis fall within the acceptable  ranges obtained in Ref.~\cite{Cordero-Cid:2018man} where a combined ATLAS and CMS Run 1 limit was used for the SM-like Higgs signal strengths.

\section{Selection of benchmark scenarios}
\label{sec:DD}

\begin{table}[b!]
\centering
\begin{tabular}{|ccc|c|c|c|}
\hline
& & & scenario B & scenario C & scenario G\\
\hline
$\lambda_{11}=0.13$ & $\lambda'_{12}=0.12$ & $-0.2<g_1<0.2$ & $\Delta_n=$50 GeV & $\Delta_n=$10 GeV & $\Delta_n=$2 GeV\\
$\lambda_{22}=0.11$ & $\lambda_{1}=0.1$ &  $-0.2<g_2<0.2$ & $\Delta_c=$60 GeV & $\Delta_c=$50 GeV & $\Delta_c=$0.8 GeV \\
$\lambda_{12}=0.12$ & & $-0.1<\lambda_{2}<0.1$ &  $\delta_c=$10 GeV & $\delta_c=$1 GeV& $\delta_c=$0.5 GeV \\
\hline
\end{tabular}
\caption{\small Input parameter values for the benchmark scenario scans in Figures \ref{ScenB}, \ref{ScenC} and \ref{ScenG}.}
\label{Tab:parscan}
\end{table} 

As discussed before, for mass-degenerate $H_1$ and $A_1$ particles to qualify as viable DM candidates, the $ZH_iA_i$ vertex, proportional to $\cos 2 \theta_h$ must vanish. Therefore, $\theta_h = \pi/4$ is the only acceptable value in the $0 \leq \theta_h < \pi$ range for the model to qualify as a viable DM framework. 

With $\theta_h = \pi/4$, the mixing between the inert doublets $\phi_1$ and $\phi_2$ is maximal and the neutral mass relations are reduced to
\bea\nonumber
&& m^2_{H_1}=
m^2_{A_1} =- \frac{1}{2} (\mu^2_1 + \mu^2_2)
+ \frac{v^2}{4} (\lambda_{31} + \lambda'_{31}+ \lambda_{23} + \lambda'_{23}  + 2\lambda_3), 
  \\[1mm]
&& m^2_{H_2}=
m^2_{A_2}=  - \frac{1}{2} (\mu^2_1 + \mu^2_2)
+ \frac{v^2}{4} (\lambda_{31} + \lambda'_{31}+ \lambda_{23} + \lambda'_{23}  - 2\lambda_3) .
\eea
The charged mass eigenstates are as presented in Eq.~(\ref{charged-masses}).
In this limit the relations for the parameters in terms of the observables also reduce to
\bea
\lambda_{23} =
\frac{1}{v^2}
\biggl(
g_1 v^2+g_2 v^2-2 {m^2_{H_1}}+2{m^2_{H^\pm_2}}
\biggr)
,
& \quad &
\lambda_{31} =
\frac{1}{v^2}
\biggl(
g_1 v^2-g_2 v^2+2 {m^2_{H^\pm_1}}-2 {m^2_{H_1}}
\biggr),
\nonumber
\\[2mm]
\lambda'_{23} =      
\frac{1}{v^2}
\biggl(
{m^2_{H_1}}-2 {m^2_{H^\pm_2}}+{m^2_{H_2}} 
\biggr)
,
&\quad &
\lambda'_{31} =  
\frac{1}{v^2}
\biggl(
{m^2_{H_1}}-2 {m^2_{H^\pm_1}}
+{m^2_{H_2}}
\biggr),
\nonumber
\\[2mm]
\mu^2_{1} =
\frac{1}{2} \left(g_1 v^2-g_2 v^2-2{m^2_{H_1}}\right)
,
& \quad &
\mu^2_{2} =
\frac{1}{2} \left(g_1 v^2+g_2 v^2-2{m^2_{H_1}}\right),
\nonumber
\\[2mm]
\lambda_{3} =
\frac{1}{v^2}
\biggl(
{m^2_{H_1}}-{m^2_{H_2}}
\biggr). &&
\label{parameters}
\eea

Taking all constraints discussed into account, we devise the following benchmark scenarios in the $\theta_h=\pi/4$ limit, using the notation
\be 
\Delta_n = m_{H_2}-m_{H_1},
\qquad
\Delta_c = m_{H^\pm_1}-m_{H_1},
\qquad
\delta_c = m_{H^\pm_2}-m_{H^\pm_1}.
\label{deltas}
\ee

In the low mass region ($ 45 \mbox{ GeV} \leq m_{\rm DM} = m_{H_1} = m_{A_1}\leq 100 \mbox{ GeV}$), using the nomenclature in accordance to the $Z_2$ symmetric I(2+1)HDM literature \cite{Keus:2014jha,Cordero-Cid:2016krd}, we define two benchmark scenarios:

\begin{itemize}
\item \textbf{B-type scenarios with $\Delta_n=50$ GeV, $\Delta_c=60$ GeV and $\delta_c=10$ GeV}
\be 
m_{H_1} = m_{A_1} \ll m_{A_2}= m_{H_2} \ll m_{H^\pm_1} \sim  m_{H^\pm_2}
\ee
where all other inert particles are much heavier than 
the mass-degenerate DM constituents $H_1$ and $A_1$ and therefore cannot coannihilate with them. 
Also, due to the absence of the $ZH_1A_1$ coupling there is no $Z$ mediated co-annihilation between $H_1$ and $A_1$. Moreover, CP-conservation forbids the coupling $h H_1 A_1$ and as a result, there are no Higgs mediated co-annihilation modes between $H_1$ and $A_1$. The only annihilation processes are through $H_1H_1 h$ and $A_1 A_1 h$ vertices.

\item 
\textbf{C-type scenarios with $\Delta_n=10$ GeV,  $\Delta_c=50$ GeV and $\delta_c=1$ GeV} 
\be 
m_{H_1} = m_{A_1} \sim m_{A_2} = m_{H_2} \ll m_{H^\pm_1} \sim  m_{H^\pm_2}
\ee
where $H_1$ and $A_1$ are close in mass with other neutral inert particles and could coannihilate through Higgs and $Z$ mediated processes.
\end{itemize}

For the heavy mass region, $m_{\rm DM} >100$ GeV, there is only one benchmark scenario that is realisable here referred to as the {G-type} one, using the nomenclature of the $Z_2$ symmetric I(2+1)HDM literature\footnote{The benchmark scenario H defined for the heavy mass region in the $Z_2$ symmetric I(2+1)HDM, with the second inert family split from the first inert family, is not realisable in the $Z_3$ symmetric I(2+1)HDM. This is due to the construction of the model which does not allow for a large mass splitting between the two charged inert states $H_1^\pm$ and $H_2^\pm$ .}.
\begin{itemize}
\item \textbf{G-type scenario with $\Delta_n=2$ GeV, $\Delta_c=0.8$ GeV and $\delta_c=0.5$ GeV}
\be 
m_{H_1} = m_{A_1} \sim m_{A_2} = m_{H_2} \sim m_{H^\pm_1} \sim m_{H^\pm_2}
\ee
where $H_1$ and $A_1$ are close in mass with all other inert particles and could coannihilate through Higgs, $Z$ and $W^\pm$ mediated processes.

\end{itemize}

\begin{figure}[t!]
\centering
\includegraphics[scale=0.3]{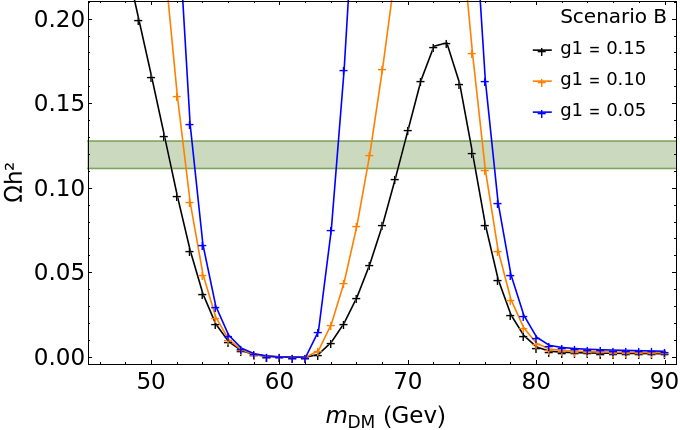}
\includegraphics[scale=0.3]{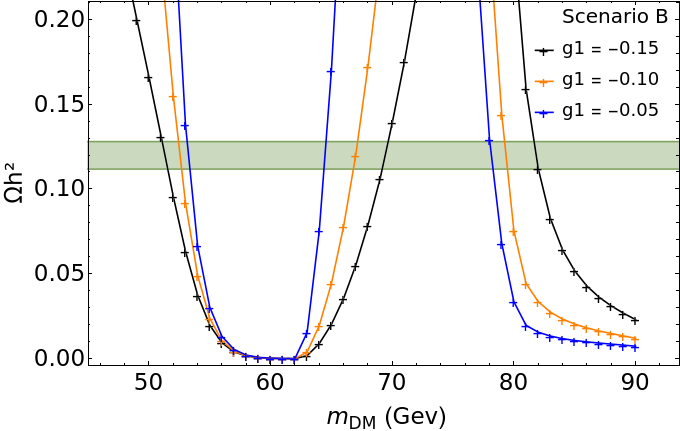}
\caption{\small The combined relic density of the DM constituents $H_1$ and $A_1$ with respect to $m_{\rm DM}$ in benchmark scenario B for varying values of positive (left) and negative (right) $g_1$ coupling. The green band represents the DM observed relic density within 3$\sigma$.}
	\label{ScenB-relic}
\end{figure}

{In Table \ref{Tab:parscan}, we show the input parameter values that we have used in our numerical analysis, which satisfy all constraints discussed in the previous section.} Our numerical analysis shows that $H_1$ and $A_1$ contribute identically to the DM relic density with identical cross sections and interactions. 
{The production and annihilation processes for the two constituents of DM, $H_1$ and $A_1$, are proportional to $H_1H_1h,A_1A_1h$ and $H_1H_1VV,A_1A_1VV$ couplings which are identical for $H_1$ and $A_1$. Therefore $H_1$ and $A_1$ contribute identically to the DM relic density with identical cross sections and interactions.}
Our analysis also confirms that varying inert self-interaction vertices (proportional to the $\lambda_1$ and $\lambda_2$ parameters),
\bea
g_{H_1A_1A_1} = -\frac{3}{2\sqrt{2}} \, (\lambda_1 + \lambda_2) \,v,
~~ && ~~
g_{H_2A_2A_2} = -\frac{3}{2\sqrt{2}} \,(\lambda_1 - \lambda_2)\,v,
\label{HHHbegin}
\\
g_{H_1A_2A_2} = \phantom{-} \frac{1}{2\sqrt{2}}\, (\lambda_1 + \lambda_2)\, v,
~~ && ~~
g_{H_2A_1A_1} = \phantom{-}\frac{1}{2\sqrt{2}}\, (\lambda_1 - \lambda_2) \,v,
\\
g_{H_1A_2A_1} = - \frac{1}{2\sqrt{2}}\, (\lambda_1 - \lambda_2)\, v,
~~ && ~~
g_{H_2A_2A_1} = -\frac{1}{2\sqrt{2}}\, (\lambda_1 + \lambda_2)\, v,
\label{HHHend}
\eea
does not affect the tree-level DM and collider phenomenology of the model.

\begin{figure}[t!]
\centering
\includegraphics[scale=0.3]{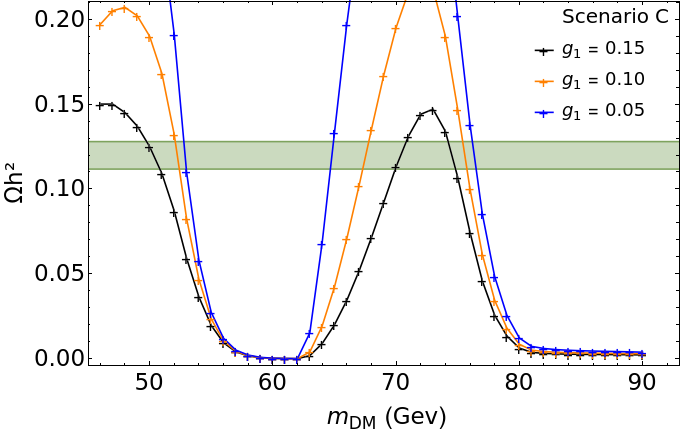}
\includegraphics[scale=0.3]{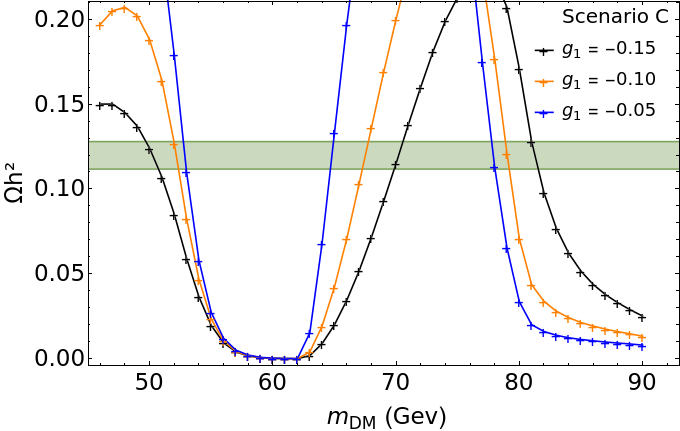}
\caption{\small The combined relic density of the DM constituents $H_1$ and $A_1$ with respect to $m_{\rm DM}$ in benchmark scenario C for varying values of positive (left) and negative (right) $g_1$ coupling. The green band represents the DM observed relic density within 3$\sigma$.}
\label{ScenC-relic}
\end{figure}

\begin{figure}[h!]
\centering
\includegraphics[scale=0.3]{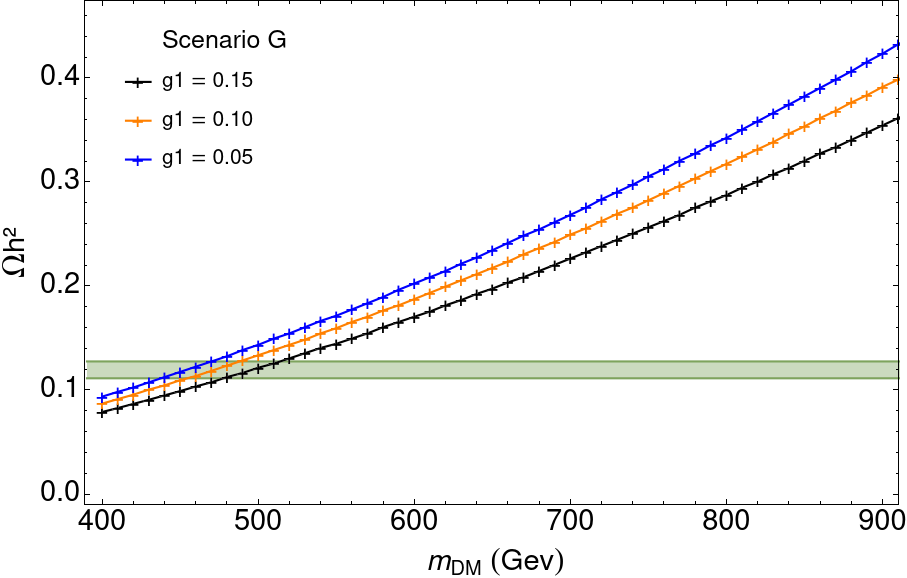}
\includegraphics[scale=0.3]{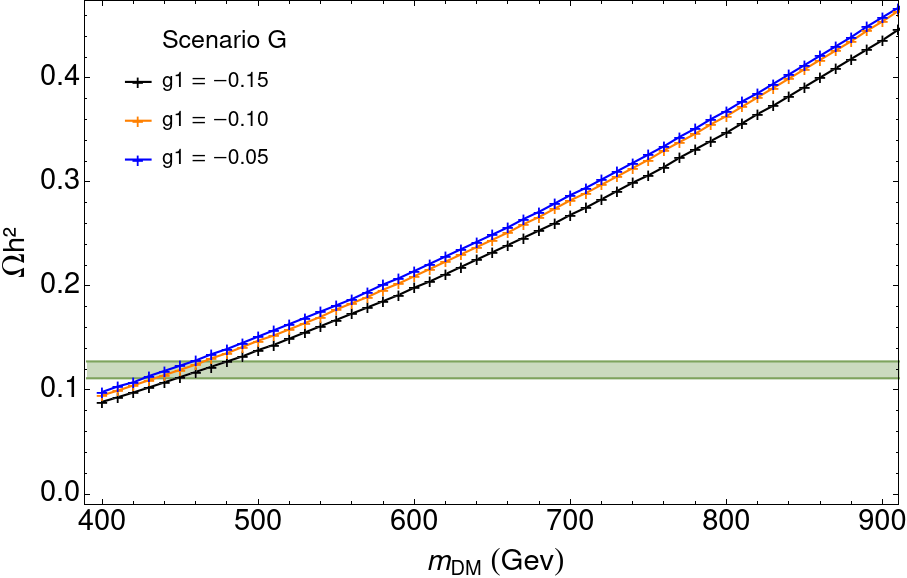}
\caption{\small The combined relic density of the DM constituents $H_1$ and $A_1$ with respect to $m_{\rm DM}$ in benchmark scenario G for varying values of positive (left) and negative (right) $g_1$ coupling. The green band represents the DM observed relic density within 3$\sigma$.}
\label{ScenG-relic}
\end{figure}

Allowing for different values of $g_1$ and $g_2$, we present in Figures \ref{ScenB-relic}, \ref{ScenC-relic} and \ref{ScenG-relic}  the combined relic density of the mass-degenerate constituents of DM, $H_1$ and $A_1$, with respect to $m_{\rm DM}$ ($m_{\rm DM}=m_{H_1}=m_{A_1}$) for benchmark scenarios B, C and G. Our analysis shows that, in all scenarios, varying $g_2$, the coefficient of the $hH_1H_2$ vertex, does not affect the behaviour of the model, while varying $g_1$, the coefficient of the $hH_1H_1$ vertex, dictates the relic density of DM. 
For scenarios B and C, for a given $g_1$, the model over produces DM for masses below 50 GeV, while DM is under-produced in the vicinity of the Higgs resonance region ($m_h/2 \approx 62$ GeV). For larger $m_{\rm DM}$, for a given $g_1$, DM is again over produced until we hit the $W^\pm$ and $Z$ resonances and DM production is suppressed. 

\section{Results} 

\subsection{DM Relic density}

As a reference value, we use the DM relic abundance measured by Planck \cite{Aghanim:2018eyx}:
\be
\Omega_{\text{\rm DM}}h^2 = 0.1198 \pm 0.0027.
\label{Planck}
\ee 
Due to the presence of two constituents of DM, $H_1$ and $A_1$, the prediction of the total relic density is given by 
\be 
\label{planck-relic}
\Omega_{\text{\rm DM}}h^2 = \Omega_{{H_1}}h^2 + \Omega_{{A_1}}h^2.
\ee 
For the numerical evaluation of the relic abundance, we use micrOMEGAs \cite{Belanger:2013oya} to show the behaviour of our benchmark scenarios B and C defined in section \ref{sec:DD}.

\begin{figure}[t]
\centering
\includegraphics[scale=1]{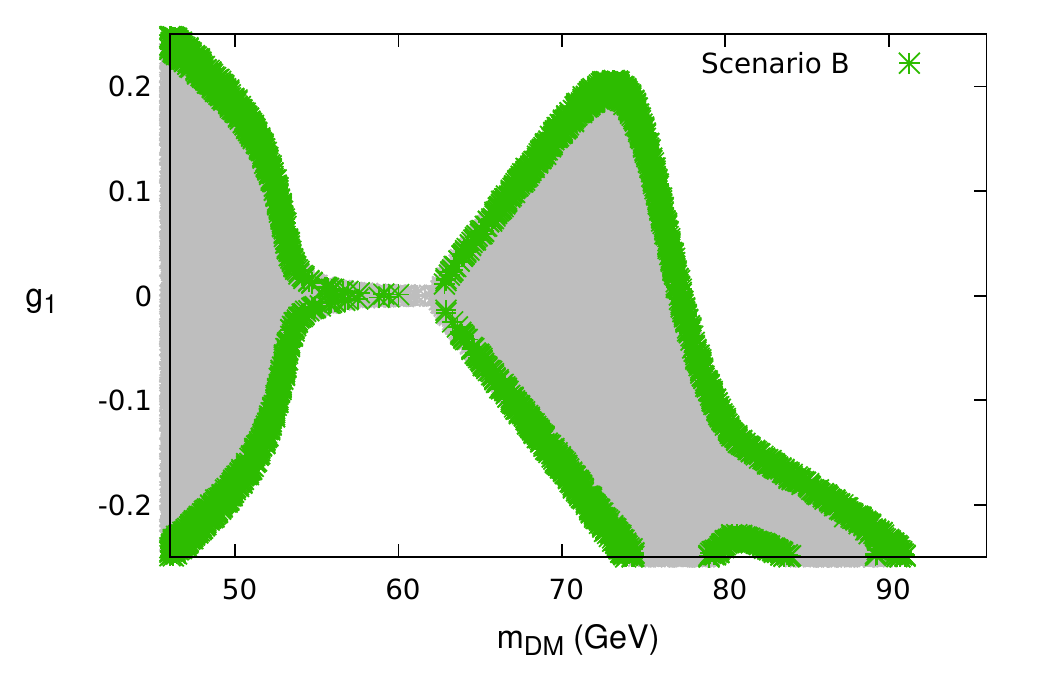}
\caption{\small Regions where the model produces the DM relic density in 3$\sigma$ agreement with Eq.~(\ref{planck-relic}) in the DM mass vs Higgs-DM coupling plane in green for scenario B for the input values in Table \ref{Tab:parscan}. The grey region represents areas where DM (co)annihilation is not strong enough and, as a result, DM is overproduced. The unshaded regions are where DM is under-produced.}
\label{ScenB}
\end{figure}

\begin{figure}[t]
\centering
\includegraphics[scale=1]{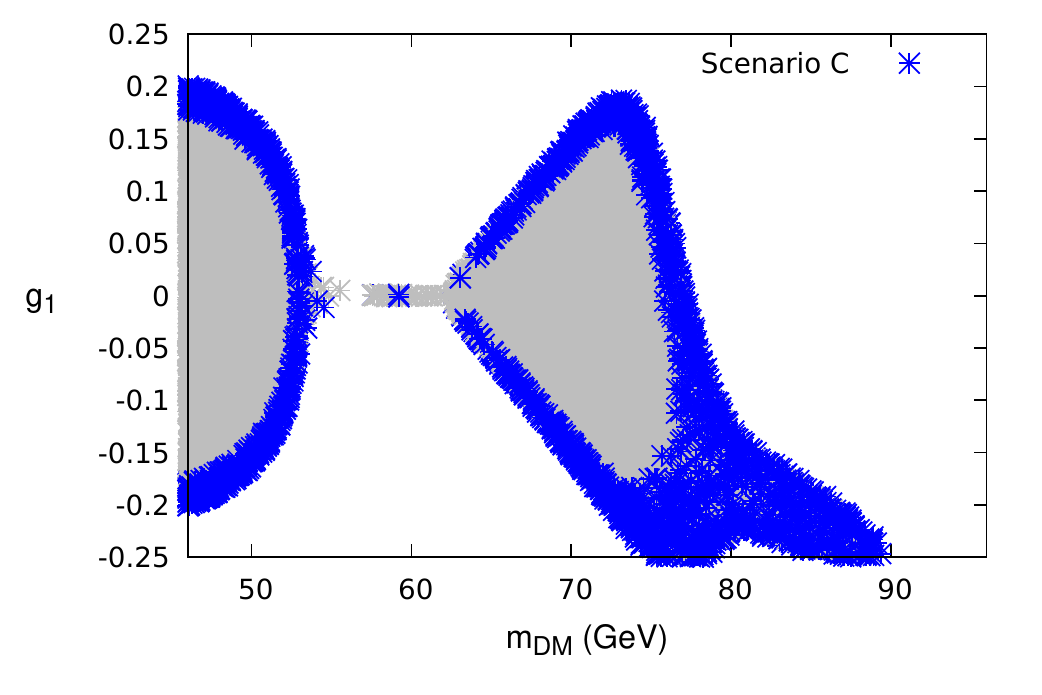}
\caption{\small Regions where the model produces the DM relic density in 3$\sigma$ agreement with Eq.~(\ref{planck-relic}) in the DM mass vs Higgs-DM coupling plane in blue for scenario C for the input values in Table \ref{Tab:parscan}. The grey region represents areas where DM (co)annihilation is not strong enough and, as a result, DM is overproduced. The unshaded regions are where DM is under-produced.}
\label{ScenC}
\end{figure}

\begin{figure}[t]
\centering
	\includegraphics[scale=1]{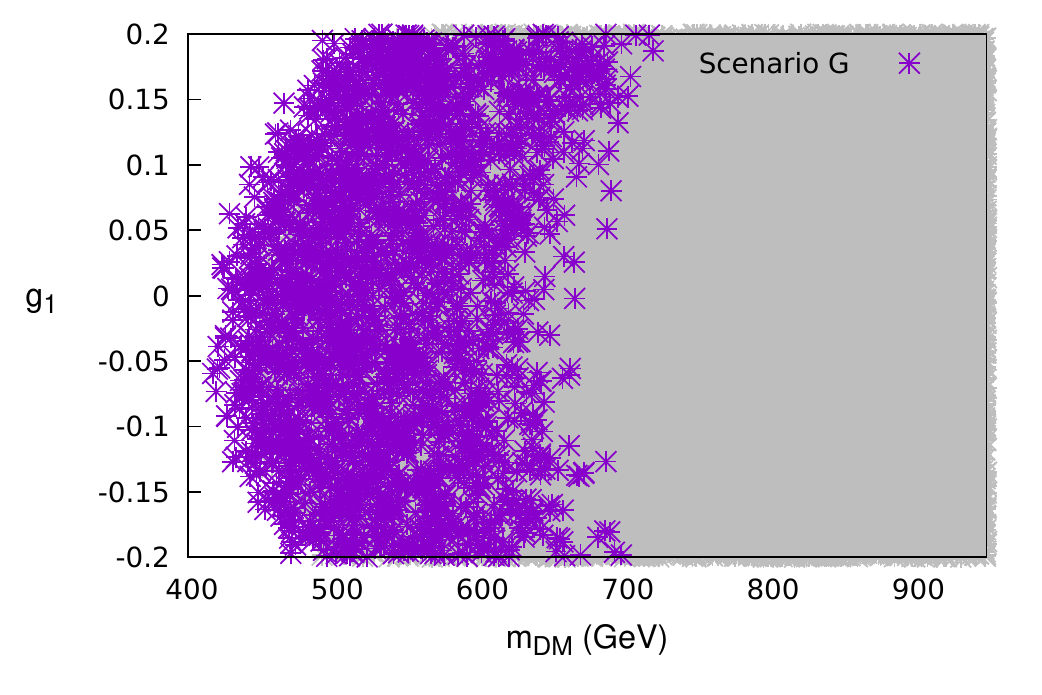}
\caption{\small Regions where the model produces the DM relic density in 3$\sigma$ agreement with Eq.~(\ref{planck-relic}) in the DM mass vs Higgs-DM coupling plane in violet for scenario G for the input values in Table \ref{Tab:parscan}. The grey region represents areas where DM (co)annihilation is not strong enough and, as a result, DM is overproduced. The unshaded regions are where DM is under-produced.}
\label{ScenG}
\end{figure}

In Figures \ref{ScenB}, \ref{ScenC} and \ref{ScenG}, we show the DM mass versus the Higgs-DM coupling and highlight the regions where the model produces the DM relic density in 3$\sigma$ agreement with Eq.~(\ref{planck-relic}). The grey regions represent areas where DM (co)annihilation is not strong enough and, as a result, DM is over produced. These regions are therefore ruled out by Planck observations. The unshaded regions are where DM is under-produced. Note that due to the presence of co-annihilation channels, in scenario C a more extensive range of Higgs-DM coupling produces the sufficient amount of DM; hence, the blue band is thicker for scenario C in Figure~\ref{ScenC} compared to the green band for scenario B in Figure~\ref{ScenB}.
In scenario G, where the charged inert scalars also coannihilate with the DM particles, the violet band showing the region with correct relic density is even broader as shown in Figure~\ref{ScenG}.

For both scenarios B and C, in the light DM mass region, the plots are symmetric for positive and negative $g_1$ values since the cross section of the Higgs mediated annihilation process $H_1H_1/A_1A_1 \to h \to f \bar f$ is proportional to $g_1^2$.
In the vicinity of the Higgs resonance region, the $H_1H_1/A_1A_1 \to h$ process is very efficient and reduces the DM abundance significantly, so much so that the Higgs-DM coupling $g_1$ has to take very small values. Recall also from Figures \ref{ScenB-relic} and \ref{ScenC-relic} that, for $m_{\rm DM} \sim m_h/2$, even negligible values of $g_1$ lead to the underproduction of DM.
The main annihilation channels in this mass region are the $H_1H_1/A_1A_1\to b\bar{b}$ process contributing $\sim$28\% and the $H_1H_1/A_1A_1\to W^{+*}W^{-*}$ process contributing $\sim$14\% of the annihilation cross section (with other annihilation channels individually sub-dominant).
As the DM mass increases, the contribution from the $b\bar{b}$ process reduces while the contribution from the $W^{+*}W^{-*}$ process grows, reaching $\sim$50\% of the total annihilation cross section for $m_{\rm DM} \sim m_{W^\pm}$.
In the heavier mass region, the point annihilation channel $H_1H_1 / A_1A_1 \to W^\pm W^\mp$ opens up and interferes destructively with the Higgs mediated process $H_1H_1/A_1A_1 \to h \to W^\pm W^\mp$.
For larger values of DM mass the point annihilation is stronger, and a larger Higgs-DM coupling is required for the effective cancellation of the $H_1H_1 / A_1A_1 \to W^\pm W^\mp$ process. This results in larger negative values of the Higgs-DM coupling in this region.
For heavier DM masses, above $m_Z$, the annihilation to gauge bosons is so strong that DM is always under-produced regardless of the value of the Higgs-DM coupling. {This is a common pattern in inert doublet models of DM, such as the I(1+2)HDM and the $Z_2$ symmetric I(2+1)HDM.}

For much heavier DM masses, where charged inert particles are also close in mass with the DM states, the destructive interference of all coannihilation processes leads to a sufficient relic density of DM, as represented by scenario G. The behaviour of the model is similar to the $Z_2$ symmetric case studied in Ref.~\cite{Keus:2015xya}, wherein, in order to get the correct relic density, larger $\Delta_c$ values require larger negative Higgs-DM couplings, smaller $\Delta_n$ values require lower DM masses and larger 
$\delta_c$ values weaken the coannihilation effects and shrink the violet band.

\subsection{Direct and indirect detection limits}
DM direct detection experiments measure the scattering of DM particles off nuclei. This interaction is mediated by the Higgs or $Z$ boson; therefore results of these experiments constrain the DM mass, as well as the Higgs-DM coupling, $g_1$, and the $ZH_1A_1$ coupling. 
{As discussed in detail before, we study a region of the parameter space where the $ZH_1A_1$ coupling is zero which is a consequence of the exact $Z_3$ symmetry of the model and the choice of $\theta_h=\pi/4$.}

\begin{figure}[h]
\centering
\includegraphics[scale=0.59]{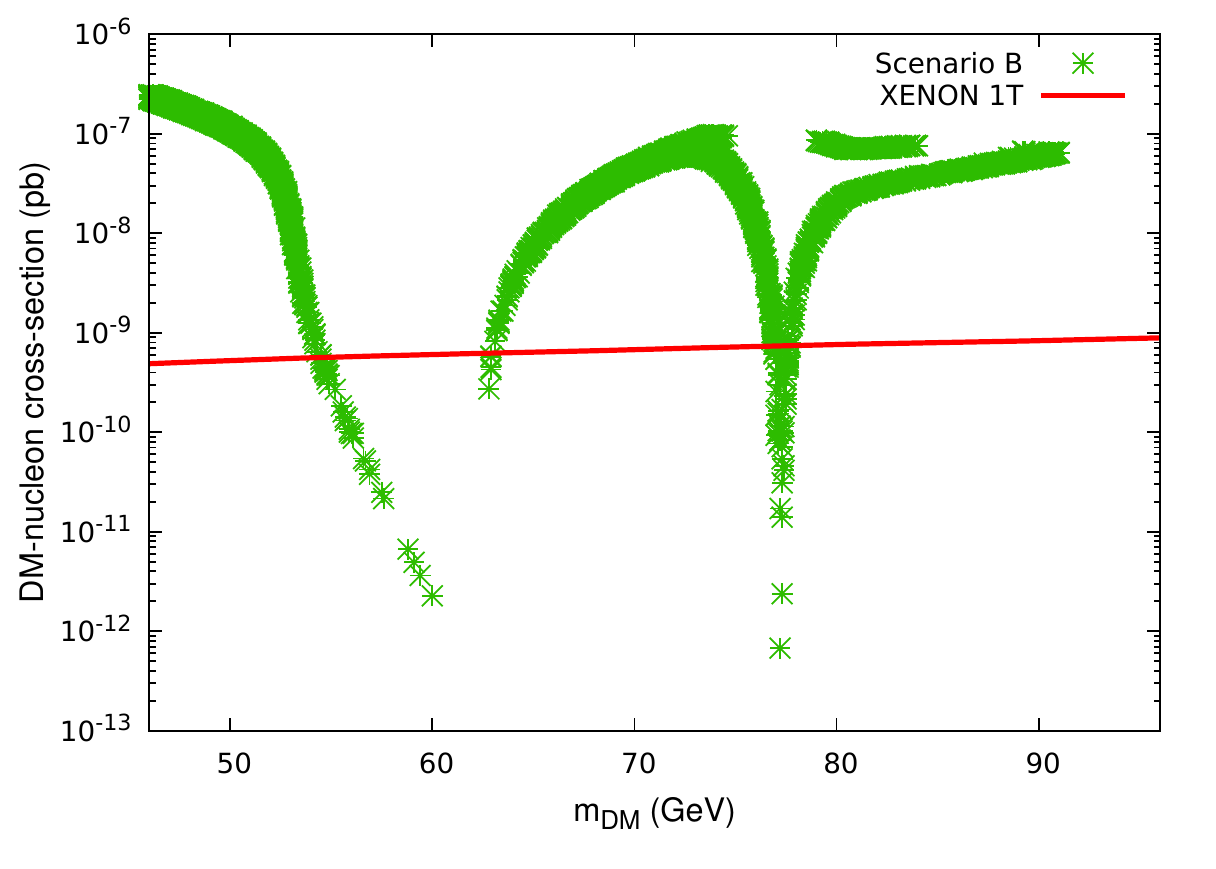}
\includegraphics[scale=0.59]{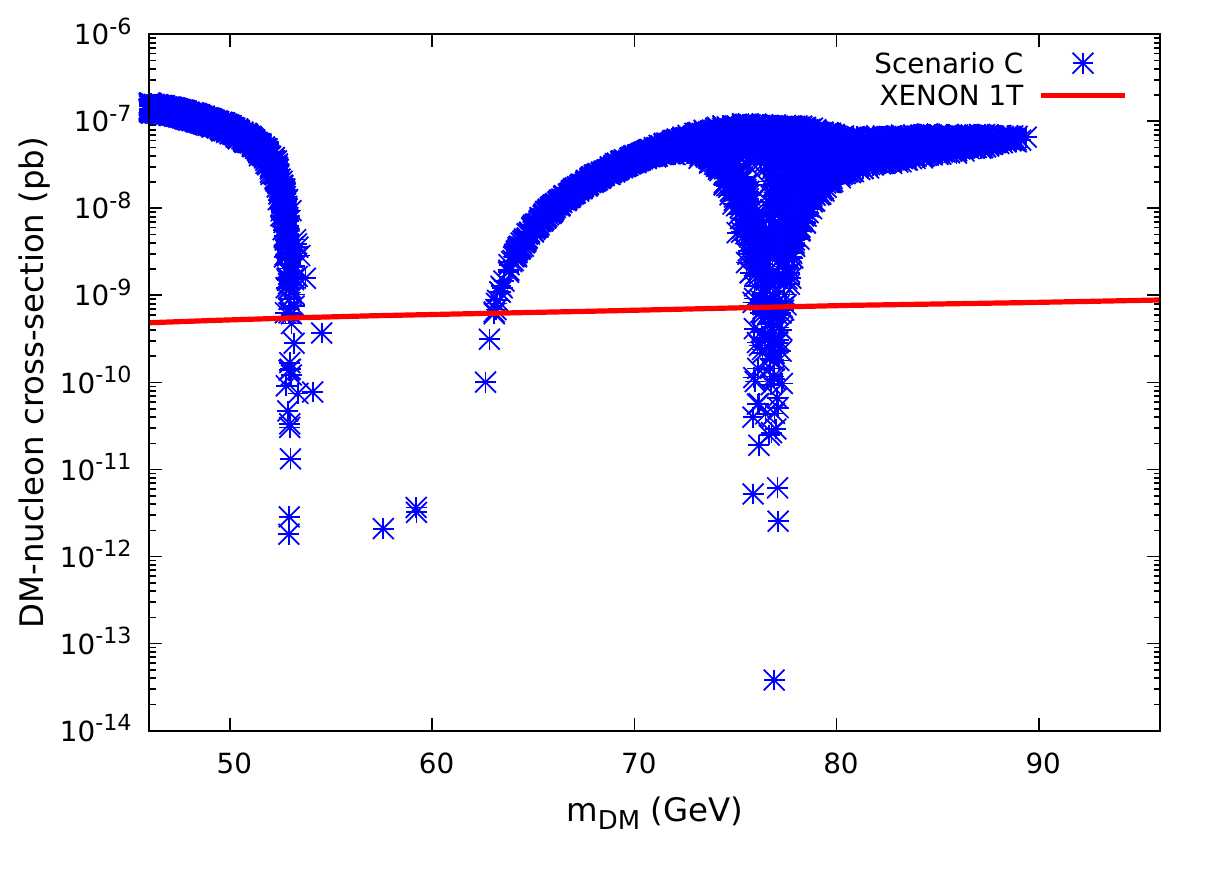}
	\includegraphics[scale=0.59]{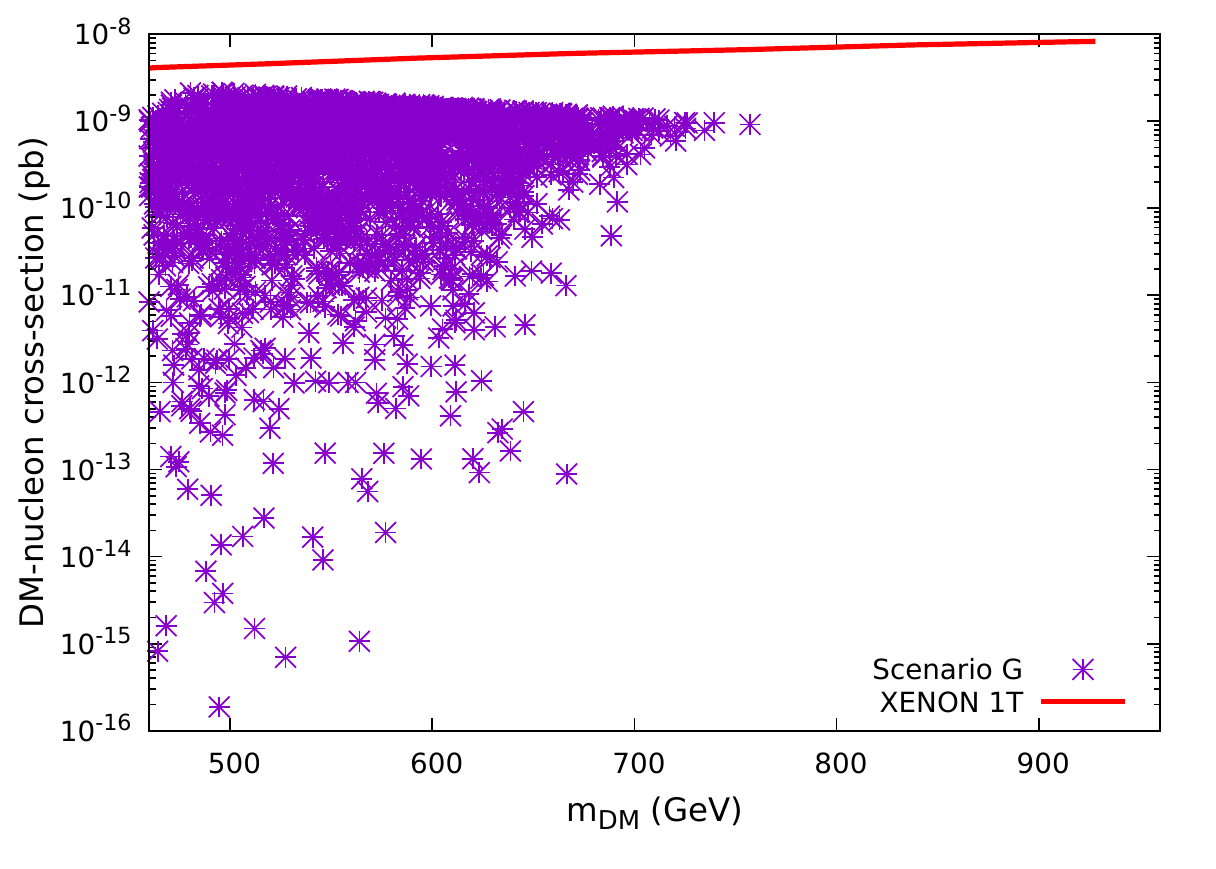}
\caption{\small Direct detection bounds on the points that saturate the relic density for scenarios B (top left), C (top right) and G (bottom). The solid red line corresponds to the current XENON1T limit above which any point is ruled out.}
\label{ScenB&C-DD}
\end{figure}

Figure \ref{ScenB&C-DD} shows the direct detection bounds on the points that saturate the relic density for scenarios B, C and G, where the solid red line corresponds to the current XENON1T limit \cite{Aprile:2018dbl}, therefore any point above this line is ruled out. 
In connection with the plots in Figures \ref{ScenB} and \ref{ScenC}, note that when the Higgs-DM coupling tends to zero, the direct detection cross section in Figure~\ref{ScenB&C-DD} is reduced to values well below the limit from XENON1T.
In the top panels, the two branches of direct detection cross section in the large mass region of the plots correspond to the two asymmetric relic density bands in Figures  \ref{ScenB} and \ref{ScenC} with different Higgs-DM coupling values.
The direct detection search by XENON1T does not exclude any relevant point in scenario G.

\begin{figure}[h]
\centering
\includegraphics[scale=0.59]{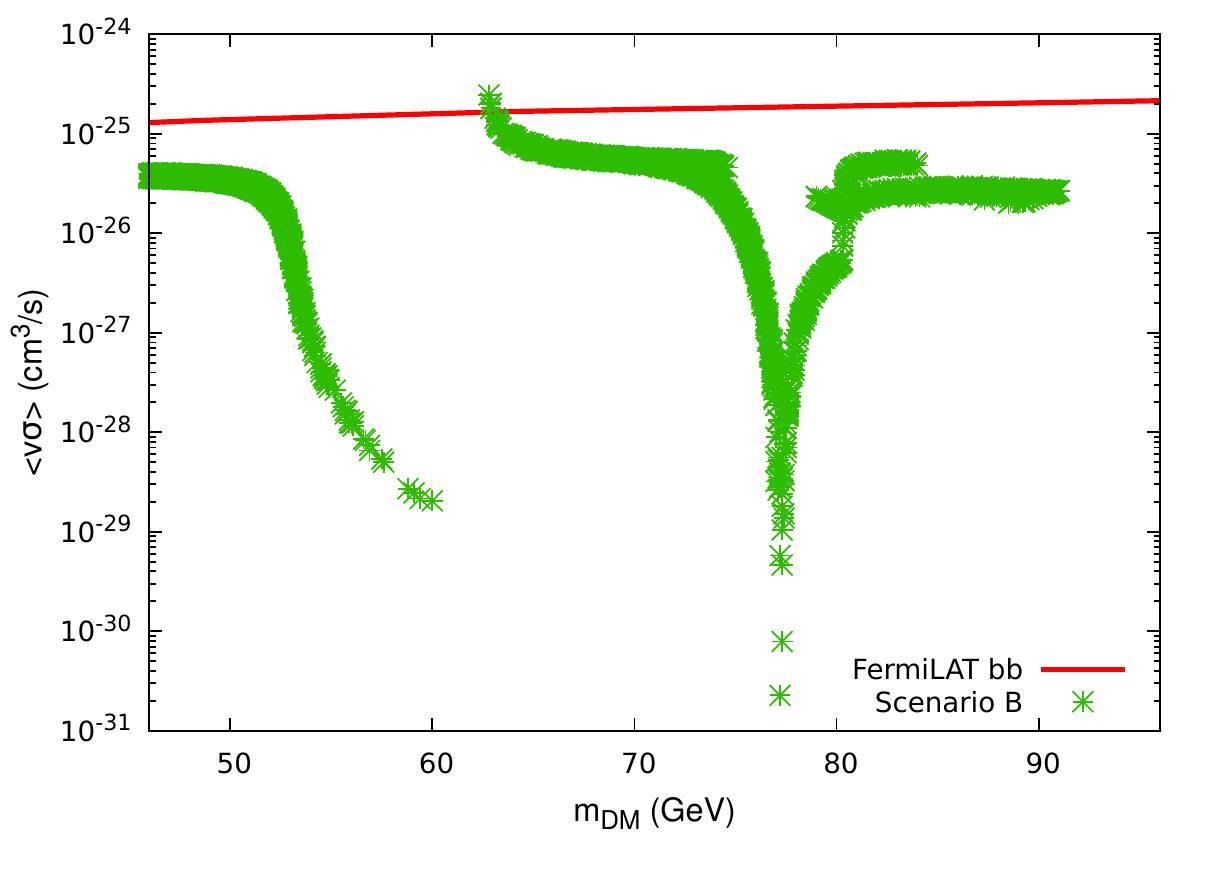}
	\includegraphics[scale=0.59]{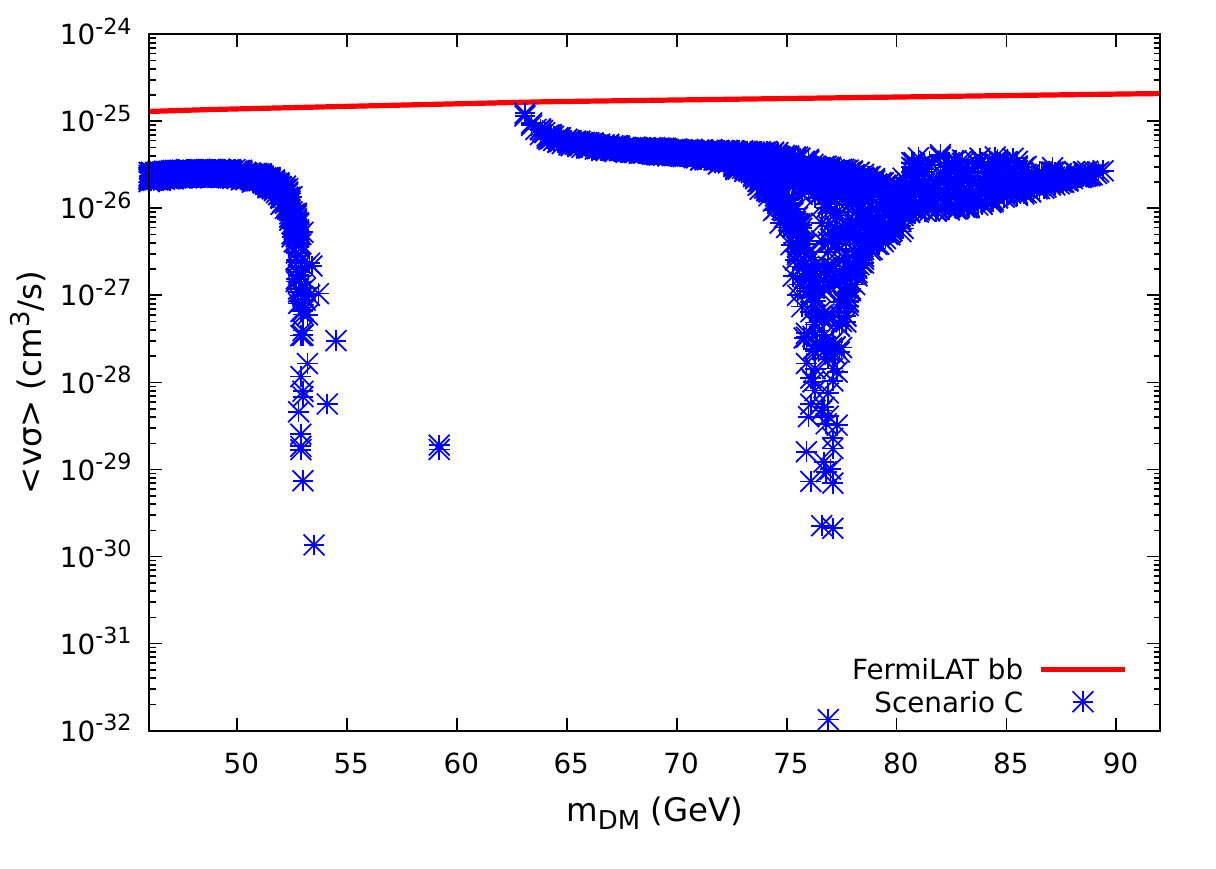}
	\includegraphics[scale=0.59]{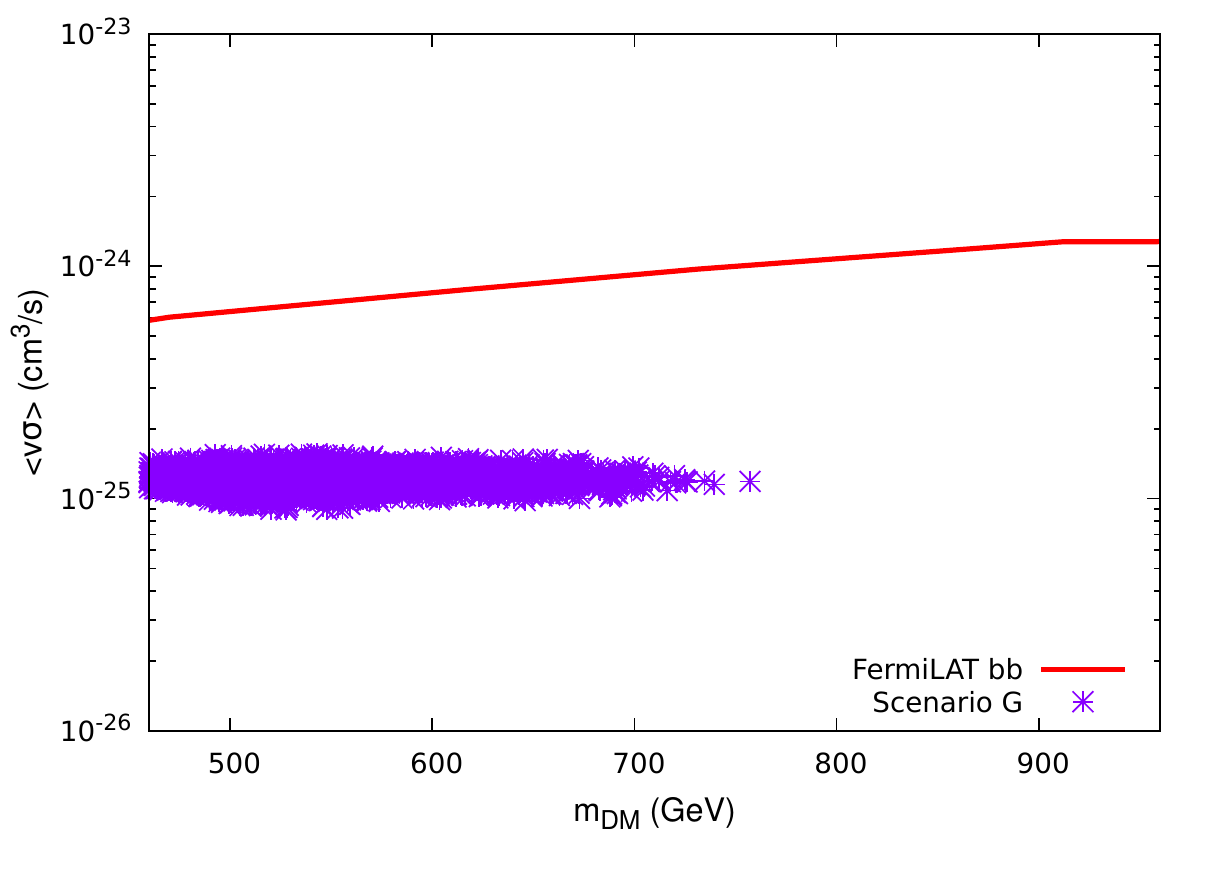}
	\caption{\small Indirect detection bounds on the points that saturate the relic density for scenarios B (top left), C (top right) and G (bottom). The solid red line corresponds to the current FermiLAT limit above which any point is ruled out.}
\label{ScenB&C-ID}
\end{figure}

Indirect detection results from FermiLAT \cite{Karwin:2016tsw} strongly constrain DM annihilation into $b \bar b$ and $\tau^+ \tau^-$.
Figure \ref{ScenB&C-ID} shows the indirect detection bounds on the points that saturate the relic density for scenarios B, C and G, where the solid red line corresponds to the current FermiLAT limit above which any point is ruled out. In the top panels, the two branches of indirect detection cross section correspond to two asymmetric relic density bands in Figures \ref{ScenB} and \ref{ScenC} with different Higgs-DM coupling values.
Notice that indirect detection bounds are much less constraining than the direct detection ones, with almost all points in scenarios B, C and G in agreement with the FermiLAT bounds.

\begin{figure}[h]
\centering
\includegraphics[scale=1]{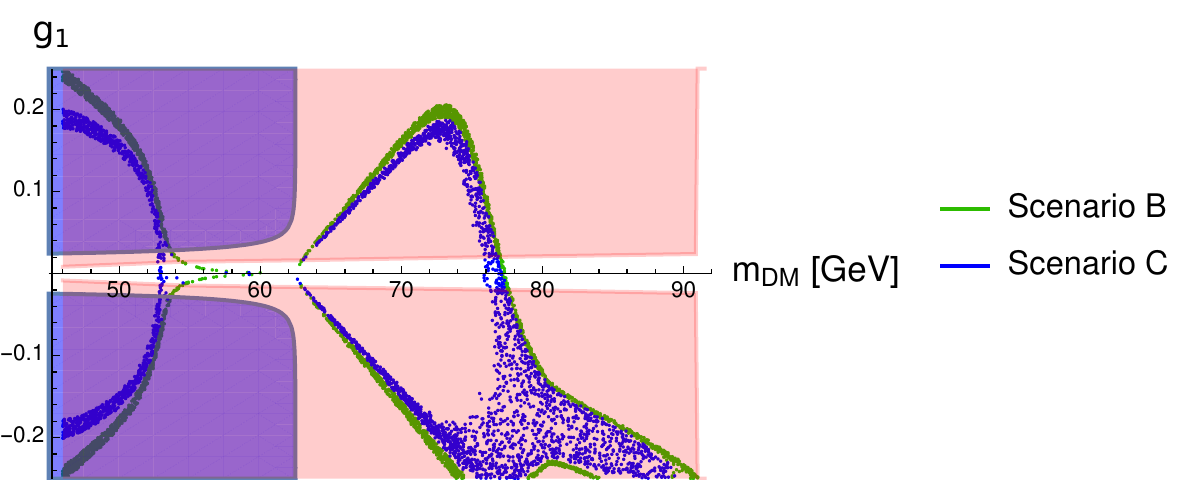}
\caption{\small The effect of the experimental constraints on the parameter space of benchmark scenarios B and C in the $m_{\text{DM}}$,$g_1$ plane. 
The red-shaded regions are excluded by direct and indirect detection experiments while blue-shaded regions are excluded by the Higgs invisible branching ration bounds.}
\label{Fig:summary}
\end{figure}

To summarise this section, in Figure \ref{Fig:summary}, we show the effect of the constraints in the [$m_{\text{DM}}$, $g_1$] plane for scenarios B and C. In red, we show the region of the parameter space excluded by direct detection bounds from XENON1T presented in Figure \ref{ScenB&C-DD}, which are more constraining that the  indirect detection bounds from FermiLAT as shown in Figure \ref{ScenB&C-ID}. The blue shaded regions are excluded by the Higgs invisible BR limits as discussed in Eqs.(\ref{Eq:Gamma_inv})--(\ref{Eq:BRinv}). The points producing the correct relic density for scenarios B and C are shown in green and blue, respectively, in $3\sigma$ agreement with Eq.(\ref{planck-relic}). In the heavy mass region, all points in scenario G leading to the correct DM relic density satisfy both direct and indirect detection bounds.

We could compare our results with those obtained in Ref. \cite{Belanger:2014bga}, wherein a $Z_3$ symmetric I(1+1)HDM plus inert singlet is explored. In their analysis,
they do not have points satisfying XENON1T limits for DM masses below 300 GeV. The latter would mean that adding a doublet to the I(1+1)HDM, rather than a singlet scalar, has the advantage of opening up regions for the low mass range in the DM mass.

\section{Hermaphrodite DM scenario}
\label{section-hermaphrodite}

As  mentioned before, the $\theta_h=\pi/4$ limit is the only viable limit for $H_1$ and $A_1$ as the two constituents of DM in the exact $Z_3$ symmetric configuration. In this case, the masses of $H_1$ and $A_1$ are degenerate and the gauge coupling $ZH_1A_1$ vanishes. In terms of  DM phenomenology, this scenario naively appears to be identical to a DM model with one complex scalar field $N=H_1+iA_i$, rather than a DM with two constituents. In fact, the relevant couplings to the annihilation of the DM particles in the early Universe and the DM scattering off nuclei are the same as those in a model with the aforementioned complex scalar DM.

However, once we turn our attention to the various parts of the Lagrangian, we realise that there are several remnants of the $Z_3$ symmetry of the model. The most interesting one is the $ZH_i^{}A_j^{} (i\neq j)$ interaction, which does not appear in a complex scalar DM model, even if the additional fields $A_2$ and $H_2$ or a complex field $N_2=H_2+iA_2$ are introduced. 
Moreover, there are significant differences in the (pseudo)scalar self-interactions. In a model with a complex scalar DM, the triple (pseudo)scalar couplings should have the form  $hN^*N = hH_1H_1+hA_1A_1$, i.e., without symmetry breaking in the dark sector. In contrast, our model has triple dark (pseudo)scalar couplings such as $H_1H_1H_1$ or $H_1A_1A_1$, as shown in Eqs.~(\ref{HHHbegin})--(\ref{HHHend}). (Further
note that our scenario does not have couplings such as $A_1A_1A_1$ and $A_1H_1H_1$ because of  CP symmetry conservation.) 
By exploring such interactions at collider experiments such as the High-Luminosity LHC 
(HL-LHC) \cite{Gianotti:2002xx} and/or a future electron-positron collider \cite{Lesiak:2019jau}, one can potentially distinguish our scenario from models with a complex scalar DM.

In summary, from the viewpoint of DM physics and collider phenomenology, one cannot identify $H_1$ and $A_1$, the two constituents of DM in our scenario, with the real and imaginary parts of one complex scalar DM particle. Indeed, while $H_1$ and $A_1$ behave identically as DM constituents, they have an opposite CP parity.  For these reasons, we coin the term Hermaphrodite DM for such a DM framework with two otherwise identical constituents but opposite CP. 

\section{Conclusions}
\label{sec:conclusions}
Motivated by two problems in the SM, from the experimental side, the absence of viable DM candidates, and, from  the theoretical side, the lack of an explanation for the three families of matter, we have postulated a 3HDM, wherein two doublets are inert (or dark), and one is active (i.e., with a SM-Higgs nature). 
This so-called I(2+1)HDM version of the 3HDM has been repeatedly studied in the literature and shown to be viable against both theoretical constraints and experimental limits when a $Z_2$ symmetry is imposed by hand onto the Lagrangian, according to which all SM fields, including the active doublet generating the $Z$, $W^\pm$ and Higgs masses measured by experiment, are even while all those emerging from the two inert doublets are odd. A consequence of this is that the lightest dark state is a viable DM candidate. 

In this paper, we have instead adopted a $Z_3$ symmetry which, combined with the $(0,0,v)$ structure for the doublet VEVs leads naturally to a novel phenomenon which we call Hermaphrodite DM here.
In this set-up, two mass-degenerate inert spin-less bosons of opposite CP, which are the lightest amongst the dark particles, contribute identically to DM phenomenology. 
We have then shown that such DM scenario is distinguishable from the complex scalar DM case.
Furthermore, all such dynamics have been obtained in the presence of known (in)direct constraints on DM, as well as those stemming from EWPOs and collider data, so that we have produced a phenomenologically successful DM framework based upon a scalar potential and VEV structure which is theoretically well-motivated. 
Finally, as a by-product of this analysis, we have also obtained that compliance with experimental results requires the DM mass to be rather light, in fact, at or below the EW scale. 
{Therefore, this ultimately opens the door to the possibility of producing peculiar signals of these dark states, such as DM clumps, core-cusp dynamics in galactic centres or 
separate cascade processes terminating with two different DM constituents at the LHC, which will  be the subject of an upcoming publication.}

\section*{Acknowledgements}
SM is supported in part through the NExT Institute and STFC Consolidated Grant ST/L000296/1. DR-C is supported by the Royal Society Newton International Fellowship NIF/R1/180813.  SM and DR-C are also partially supported by the H2020-MSCA-RISE-2014 grant no. 645722 (NonMinimalHiggs).
TS and SM are partially supported by the Kogakuin University Grant for the project research ``Phenomenological study of new physics models with extended Higgs sector''.
AA acknowledges support from CONACYT project CB-2015-01/257655 and SNI (M\'exico).
VK acknowledges financial support from Academy of Finland projects ``Particle cosmology and gravitational waves'' no. 320123
and ``Particle cosmology beyond the Standard Model'' no. 310130. 
JH-S has been supported by SNI-CONACYT (Mexico), VIEP-BUAP and PRODEP-SEP (Mexico) under the grant ``Red Tematica: Fisica del Higgs y del Sabor''. 
The authors acknowledge the use of the IRIDIS High Performance Computing Facility, and associated support services at the University of Southampton, in the completion of this work.

\end{document}